\documentclass{aa}                                                      

\usepackage{graphicx}

\begin{document}

\newcommand{\msol}{\rm M_{\sun}}
\newcommand{\Mdot}{\raisebox{0.25ex}{$\stackrel{.}{M}$}}
\newcommand{\pai}{Paper\,\,I}
\newcommand{\paii}{Paper\,\,II}

\title{The Cygnus X region} 
\subtitle{XXII. A probable HAeBe star with a giant bipolar outflow
in DR 16}
\author{Otto P. Behre\inst{1}
        \and Heinrich J. Wendker\inst{1} 
        \and Lloyd A. Higgs\inst{2} 
        \and Thomas L. Landecker\inst{2}}
\offprints{H.J.Wendker, \\
           \email {hjwendker@hs.uni-hamburg.de}}
\institute{Hamburger Sternwarte, Gojenbergsweg 112, D-21029 Hamburg,
           Germany 
           \and National Research Council, Herzberg Institute of
           Astrophysics, Dominion Radio Astrophysical Observatory, 
           Box 248, Penticton, B.C. V2A 6J9, Canada}
\date{Received 22 September 2003/ Accepted 03 November 2003}

\abstract{
From medium-resolution radio images, DR 16 was suspected to be a large
cometary nebula. To test this suggestion we obtained a higher
resolution (15\arcsec) VLA continuum map. We also analyzed data from
the Canadian Galactic Plane Survey in continuum, \ion{H}{i} line, and
IR.  These data were supplemented by published near-infrared (J, H, K)
stellar photometric results and MSX 8.28 $\mu $m data. We suggest that
DR 16 is the diffuse \ion{H}{ii} region of an ongoing star formation
site at a distance of about 3 kpc. The complicated radio picture
arises from the superposition of diffuse \ion{H}{ii} with the remains
of a giant bipolar outflow. The outflow was generated by a probable 
Herbig AeBe star, and the lobes are the remnants of its working
surfaces. Additional ring-like features are discussed.  DR 16 is part
of a larger volume of space in the local spiral arm where star
formation is an ongoing process.
\vspace{-0.3cm}
}
\titlerunning{
A Herbig Ae/Be star with a giant bipolar outflow
in DR 16}
\authorrunning{O.P.Behre et al.}

\maketitle

\section{Introduction}

In the direction of Cygnus X our line of sight is tangential to the
local spiral arm, and the superposition of many objects makes the
region one of the most complex in the sky.  Although the radio
emission in \object{Cyg X} is mainly thermal in origin, individual
sources reveal a surprisingly rich spectrum of properties, with
objects in different stages of evolution occupying volumes of
space with very different star formation histories.  Uncovering the
nature of individual objects and establishing their distances demands
extensive and thorough observations. 
\begin{figure}[h]
\resizebox{\hsize}{!}{\includegraphics{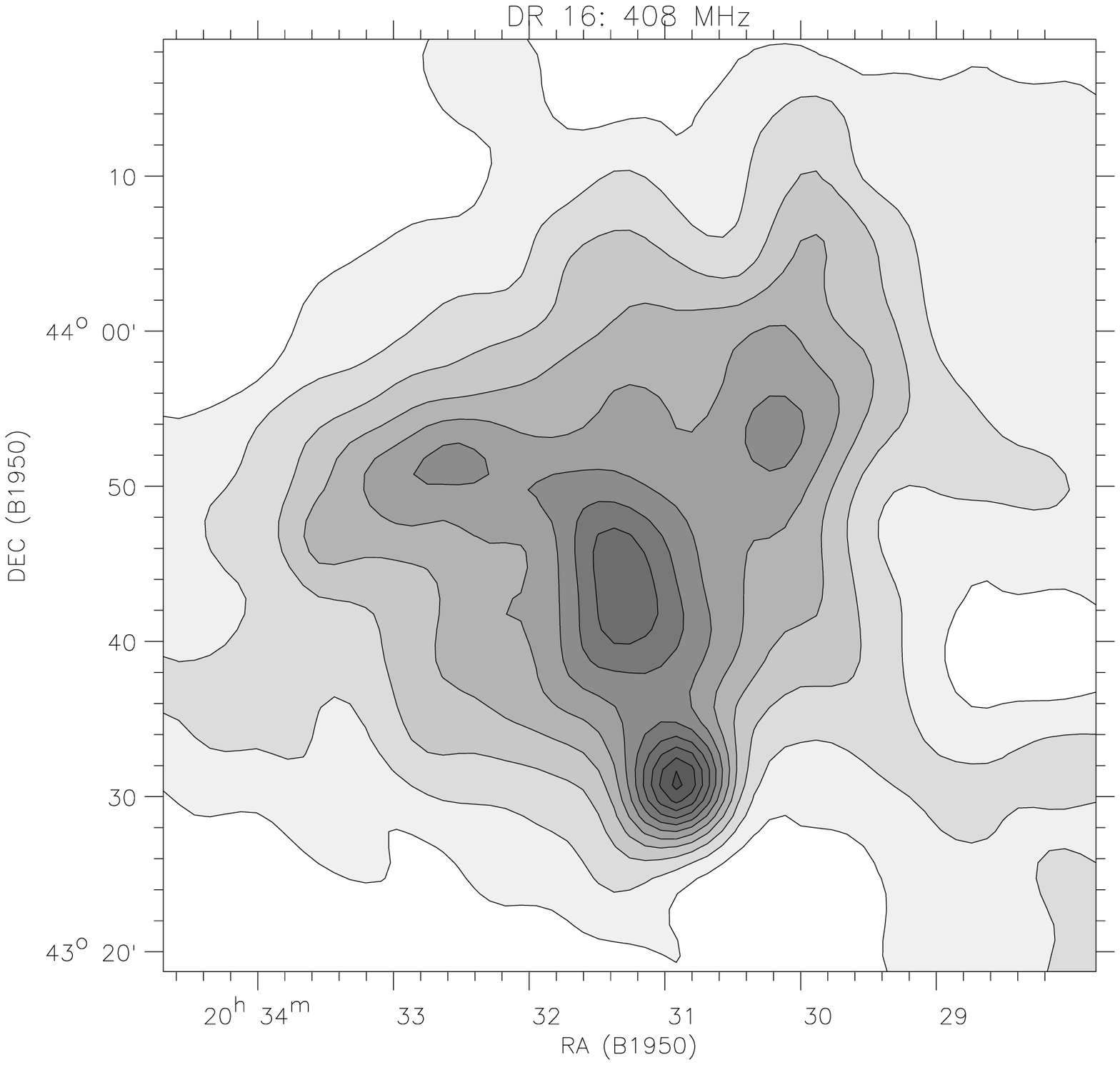}}
\caption{408 MHz map of the DR 16 area extracted and enlarged from
the map in Paper XVIII. The coordinates are B1950.
The contours are in steps of 40 K $T_\mathrm{b}$ beginning at 140 K.}
\label{kar74}
\end{figure}
The most recent radio continuum
surveys (Wendker et al. \cite{cx18}, Paper XVIII) showed that a
resolution of about 1\arcmin\ begins to resolve many sources, so that
observations of higher resolution will be effective in imaging their
structure and uncovering their true nature. We have already studied
the peculiar supernova remnant \object{G76.9+1.0} (Landecker et
al. \cite{tll1}) and the ring \object{G79.29+0.46} around a suspected
LBV (Higgs et al. \cite{lah2}).  
\begin{figure*}
\resizebox{\hsize}{!}{\includegraphics{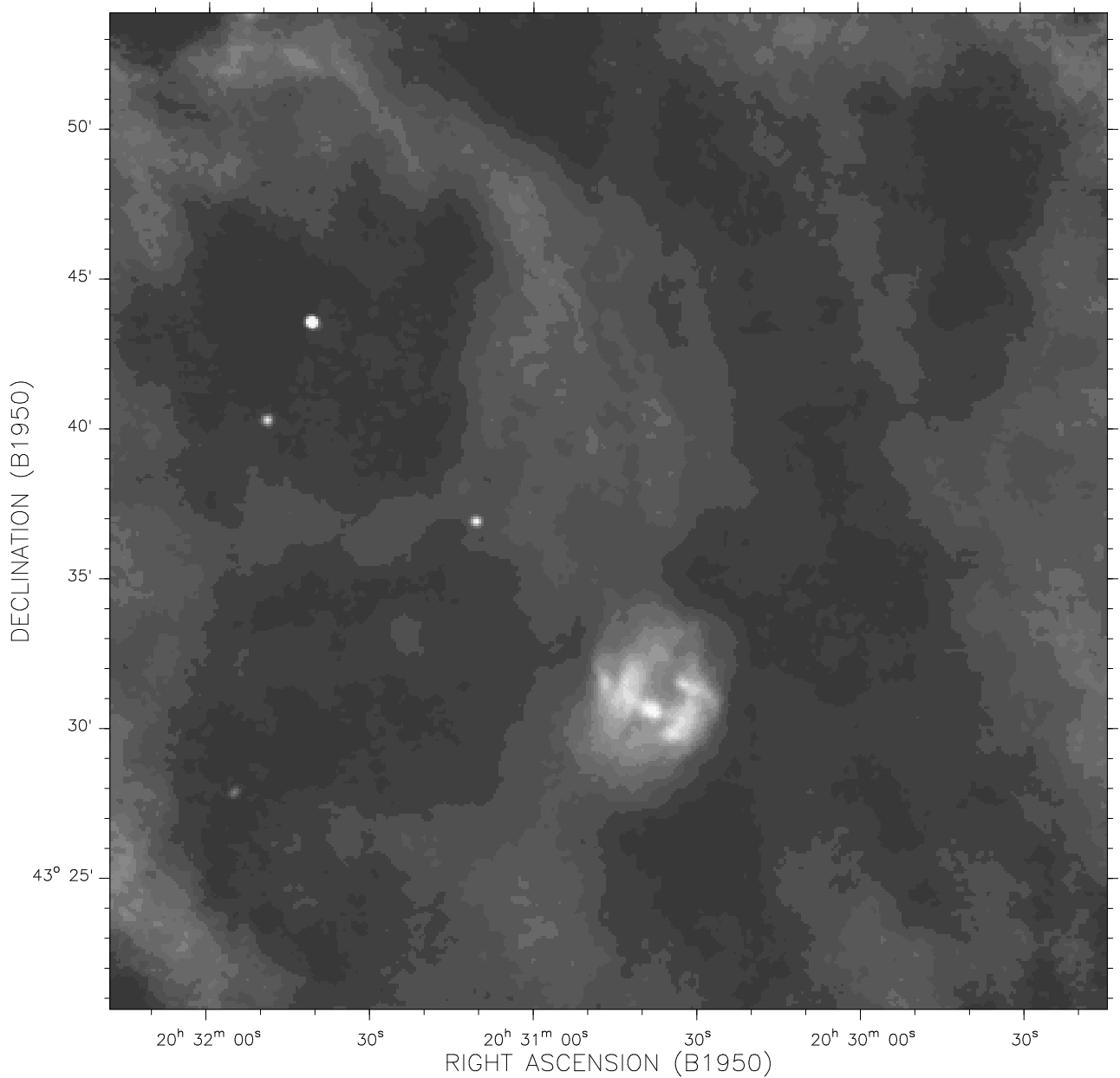}}
\caption{VLA map of DR 16 at 1515 MHz. The gray scale is non-linear in 
order to emphasize the large-scale structure of the fainter emission
and brightness increases from black to white.}
\label{vlaganz}
\end{figure*}
Another interesting feature is
\object{DR 16} (Downes \& Rinehart \cite{dr}) -- \object{G82.0+2.3} --
and its surroundings.  Images at 408 MHz (Paper XVIII) and 4.8 GHz
(Wendker \cite{cx15}, Paper XV) showed a cometary
nebula. Recombination line observations at 6 cm (Piepenbrink \&
Wendker \cite{cx17}, Paper XVII) gave nearly the same radial velocity
at three positions, suggesting that the feature is one physical
object, with a pronounced head and an impressive flaring tail about
$0.5^{\circ}$ in length. Fig.~\ref{kar74} clearly conveys this
impression. Nevertheless, in Paper XVIII we labelled the feature as a
ridge (\object{CXR 10}).  Optically only the northern part of the tail
is visible in H$\alpha $ (filaments \object{DWB 155} and \object{DWB
157} in the catalogue of Dickel et al. (\cite{cx5}, Paper V)). Visual
extinction in front of the head rises to more than $5^\mathrm{m}$
(Dickel \& Wendker \cite{cx11}, Paper XI), totally obscuring it.

The physical picture commonly associated with the term `cometary
nebula' is an object moving through its surroundings and producing
some sort of wake. Supersonic motion produces a bowshock object, and
is not implied in this case. Since sub-sonic or mildly supersonic
velocities (less than 10 km/s) do not produce large extensions, a tail
of length $0.5^{\circ}$ immediately implies a very nearby object, or
an exceptionally large one.

In this paper we present a new VLA 20-cm image with resolution
15\arcsec\, which reveals that the apparent head of the comet (which
we now label DR 16) has a dumbbell structure. Furthermore, little
evidence remains of a connection between the head and the other parts
of the ridge (which we now refer to as the tail). We also present new
observations from the database of the Canadian Galactic Plane Survey
(the CGPS; see Taylor et al. \cite{art}) and collect other published
observations for a comprehensive study of the source. In Sect. 2 to 5
we describe new observations at radio and infrared wavelengths and
present the results. A distance is suggested in Sect. 6. In Sect. 7 to
9 we discuss the source properties and suggest a source model and a
source history.

\section{Observations}

Inspection of Fig.~\ref{kar74} showed that new observations with
resolution considerably better than 1\arcmin\ would be needed to
resolve the head of the apparent cometary nebula, with a substantial
field of view needed to reveal the complete picture. In this section
we will describe targeted VLA observations and the new CGPS data
products at 1420 MHz in continuum and 21-cm line. We also summarize
results from other publicly available datasets.

\subsection{A VLA map at 1.5 GHz}

The observations at 1.5149 GHz were obtained with the NRAO Very Large
Array (VLA)\footnote{The U.S. National Radio Astronomy Observatory is
operated by Associated Universities Inc., under contract with the
National Science Foundation.} in the C/D and D configurations.
Concurrent observations of other sources of interest in Cyg X have
been described by Landecker et al. \cite{tll1} and Higgs et
al. \cite{lah2}. Instrumental parameters are summarized in Table
~\ref{tab-vla}.
\begin{table}[h]
\caption{\label{tab-vla} Parameters of VLA observations.}
\begin{tabular}{ll}
\hline
Field centre: & \\
$\alpha $(B1950) & $20^h\,30^m\,45^s$ \\
$\delta $(B1950) & $43\degr \,37\arcmin $ \\
Frequency & 1.5149\,\,GHz\\
Date C/D configuration & 1988 May 17 \\
Date D configuration &  1988 August 22 \\
On-source time C/D & 39.7 min \\
On-source time D & 19.7 min \\
Calibrators: & \\
~~flux density & 3C 286; ~~14.39 Jy \\
~~phase & 2050+364 \\
~~polarization & 3C 138 \\
Synthesized beam & 15\arcsec $\times $ 15\arcsec \\
Field of view (HPW) & 29.22\arcmin \\
Synthesized map size & 21.25\arcmin\ $\times$ 21.25\arcmin\ \\
Surface brightness conversion & 1 Jy/beam $\leftrightarrow $ 
2368 K \\
\hline
\end{tabular}
\end{table}
\begin{table}[h]
\caption{\label{tab-drao} Parameters of the DRAO CGPS data.}
\begin{tabular}{ll}
\hline
CGPS mosaic MN2 & (release April 2002) \\
Centre frequency & 1.420\,\,GHz\\
Continuum: & Stokes I\\
Synthesized beam & 71.5\arcsec $\times $ 49.1\arcsec ,\,PA -36.2\degr \\
rms noise & 40 mK \\
Line synthesized beam & 85.4\arcsec $\times $ 58.6\arcsec ,\,PA -36.0\degr \\
rms noise & 2.5 K \\
Radial velocity coverage & -164 $\rightarrow $ +58 km/s\\
Channel spacing & 0.82446 km/s \\
Velocity resolution & 1.32 km/s\\
\hline
\end{tabular}
\end{table}
The phase centre was not set on the maximum of the head but several
arcmin to the north in order to achieve better coverage of the whole
object.  The two VLA configurations were combined and CLEANed using
standard procedures. The map units were expressed in brightness
temperature T$_\mathrm{b}$. We used the polar diagram corrected map as
several components appeared to be too extended to allow a simple
post-integration factor to be applied. 
No short-spacing information was added. First, the head, with its size
of $\sim$5\arcmin\ was adequately covered. Second, the overall
structure is better depicted in the CGPS image discussed below.

Fig.~\ref{vlaganz} shows a gray scale image of the whole field, with a
grayscale chosen to emphasize the fainter large-scale structure. The
overall impression of a cometary nebula has been lost, because the
`western ear' has become detached from the rest of the tail and the
`eastern ear' seems to be part of a long curved ridge.  
\begin{figure*}
\resizebox{\hsize}{!}{\includegraphics{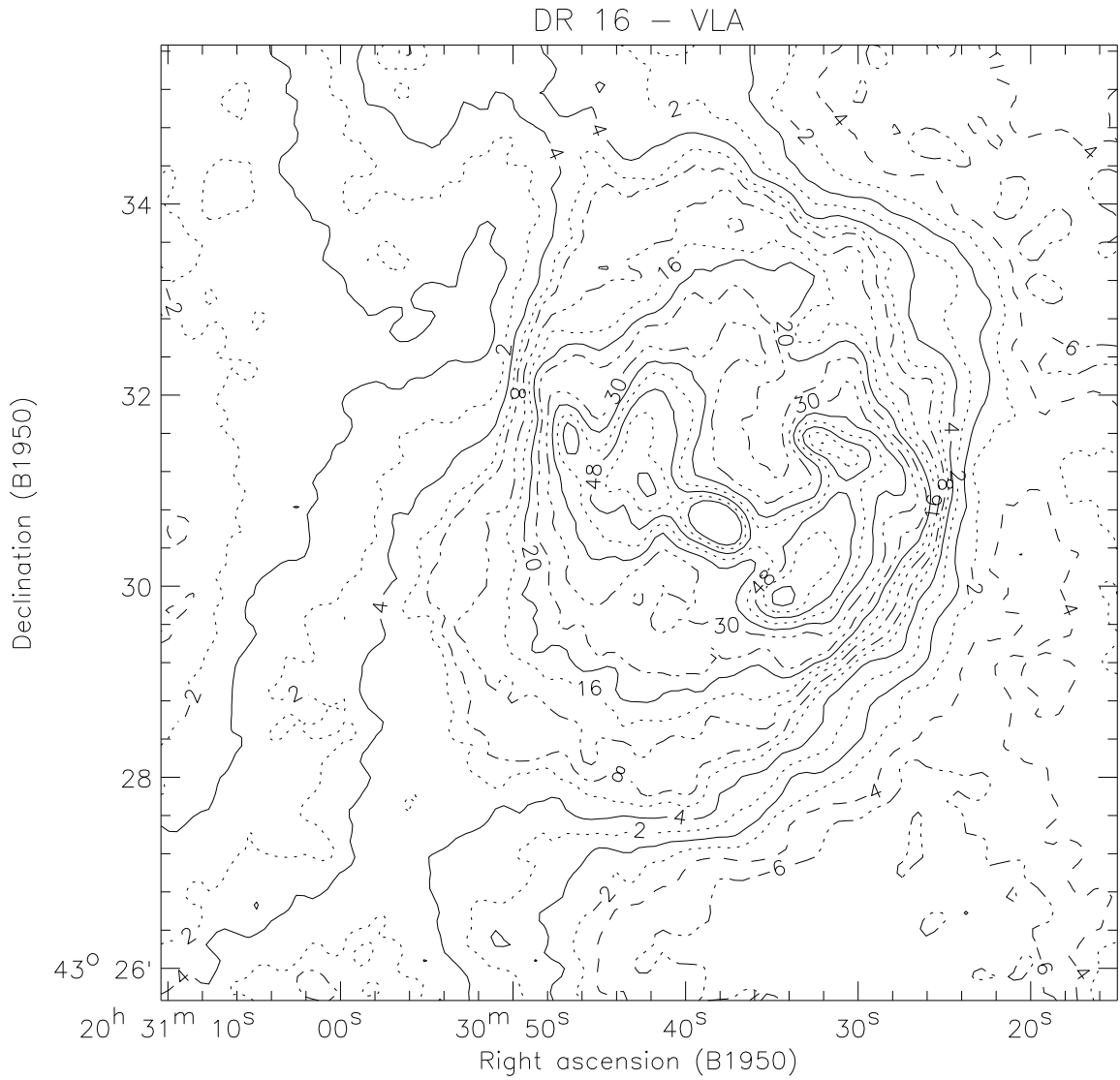}}
\caption{Enlargement of the DR 16 area proper from Fig.~\ref{vlaganz}.
The contours are in $K\,T_b$. The contours are in steps of 2 K up to 8
K, in steps of 4 K up to 24 K, and in steps of 6 K up to 60 K.
The solid, dotted or dash-dotted contours are assigned in order to
enhance clarity.
}
\label{vladr16} 
\end{figure*} 
\begin{figure*}
\resizebox{\hsize}{!}{\includegraphics{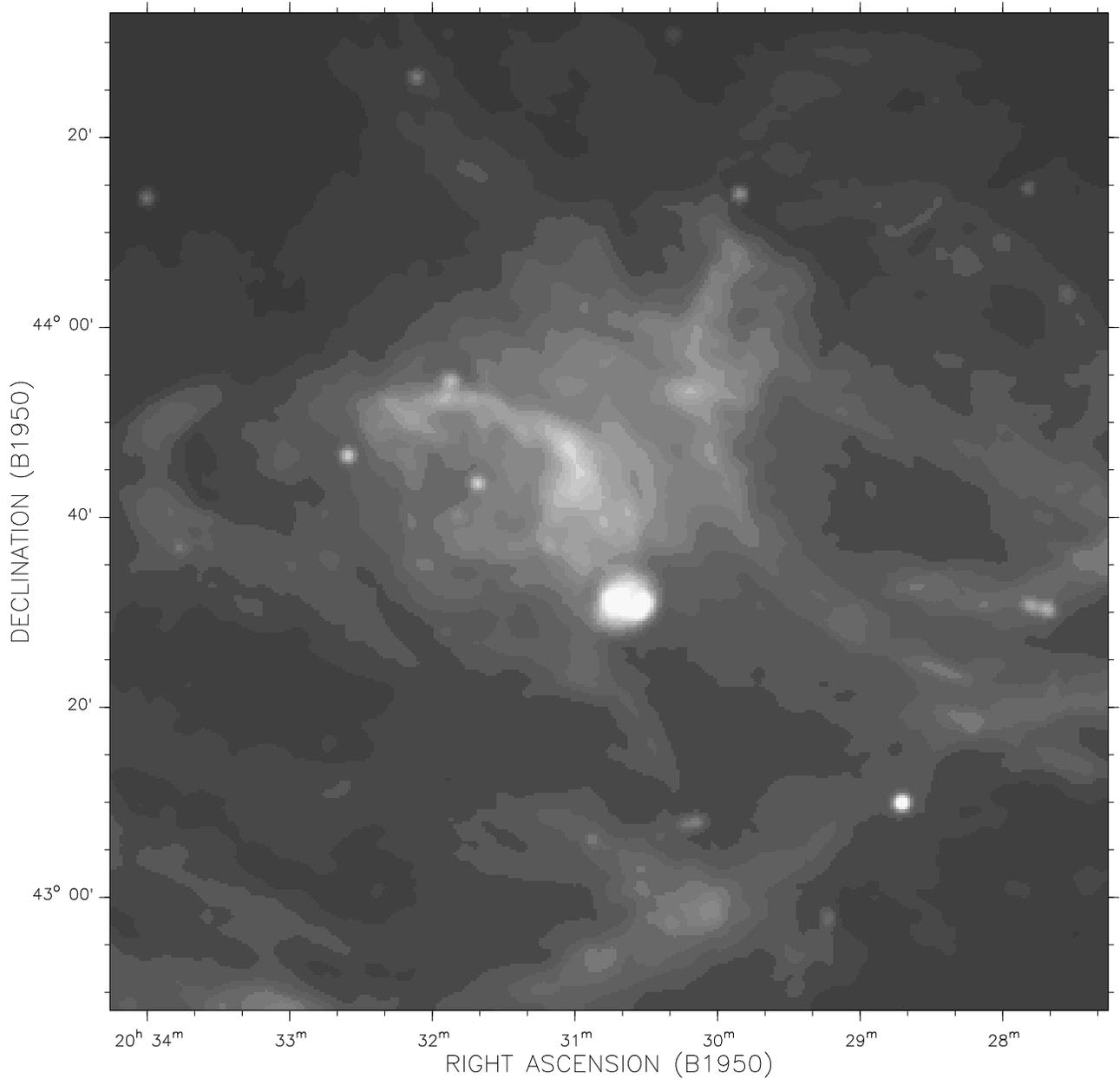}}
\caption{CGPS 1420 MHz grayscale image of DR 16 and surrounding 
region. The 
grayscale is non-linear, chosen to emphasize the large-scale structure 
of the fainter emission. Brightness increases from black to white. }
\label{cgpscont21} 
\end{figure*} 
\begin{table*}
\caption{\label{tab-other} Other data sets used in this paper.}
\begin{tabular}{llll}
\hline
Frequency or & Telescope & Angular    & Reference \\
wavelength   &           & resolution &            \\
\hline
151 MHz      & CLFST & 102\arcsec $\times $ 70\arcsec , PA 0\degr & 
Vessey \& Green \cite{vg1} \\
327 MHz      & WSRT & 78\arcsec $\times $ 54\arcsec , PA 0\degr &
Rengelink et al. \cite{rtb} \\
408 MHz      & DRAO SRT & 312\arcsec $\times $ 210\arcsec , PA 0\degr &
Paper XVIII \\
4800 MHz     & MPIfR 100 m-RT & 161\arcsec $\times $ 153\arcsec , PA 0\degr &
Paper XV \\
100 $\mu $m  & IRAS-HIRES & 132\arcsec $\times $ 102\arcsec & 
Cao et al. \cite{caoiras} \\
60 $\mu $m   & IRAS-HIRES & 102\arcsec $\times $ 60\arcsec & 
Cao et al. \cite{caoiras} \\
25 $\mu $m   & IRAS-HIRES & 66\arcsec $\times $ 34\arcsec & 
Kerton \& Martin \cite{keriras} \\
12 $\mu $m   & IRAS-HIRES & 67\arcsec $\times $ 33\arcsec & 
Kerton \& Martin \cite{keriras} \\
8.28 $\mu $m & MSX  & 20\arcsec $\times $ 20\arcsec & 
Price \cite{price}, IRSA \\
\hline
\end{tabular}
\end{table*}
The head of the cometary nebula shows complex sub-structure, as seen
in the detailed contour map of Fig.~\ref{vladr16}. Although the outer
contours are nearly circular, with an overall size of $\sim$5\arcmin\,
the source has a dumbbell shape with some superimposed small-scale
features. DR 16 is apparently placed asymmetrically on the tail,
which runs north-south. These complex gradients make it difficult to
separate DR 16 and its components from the background structure.

\subsection{CGPS 21 cm continuum and line}

The CGPS observations include the Cyg X region. One of the outstanding
properties of the Survey is the coverage of a large part of the
northern Galactic hemisphere at a number of wavelengths with a
resolution of around 1\arcmin\ or slightly worse, covering the radio
and infrared regimes\footnote{The CGPS data are available to the
astronomical community at http://www.cadc.hia.nrc.gc.ca}.  Its
``backbone'' is provided by the 21-cm continuum and line observations
from the Synthesis Telescope of the Dominion Radio Astrophysical
Observatory (DRAO). A description is given by Taylor (\cite{art}), and
here we present only those parameters relevant to the DR 16 area (in
Table ~\ref{tab-drao}). The mosaic has short-spacing information
added, with continuum data taken from the Effelsberg survey of Reich
et al. (\cite{rrf}), and line data from the DRAO 26-m Telescope survey
of Higgs \& Tapping (\cite{ht1}).


The CGPS continuum image of the area is presented in
Fig.~\ref{cgpscont21}, intended to show the very complex environs of
the alleged cometary nebula by using the scaling to emphasize faint
emission. The area is criss-crossed by filamentary structures, which
could be anything from fragments of shock fronts to small bubbles:
none is yet identified or explained. The `western ear' is now seen as
a brightness enhancement on a filament apparently unconnected with DR
16, and should be discussed in another context. Such a conclusion is
not immediately possible for the `eastern ear', but it seems
unavoidable that we must abandon the picture of a cometary nebula. In
this image DR 16 appears as a fairly compact source. Whether the
`tail' is physically connected to DR 16 remains an open question, to
which we will return later.

The CGPS 21-cm line data reveal a bewildering array of structure
around DR 16. However, two points emerge. First, there are no emission
structures that can be easily related to DR 16. Second, we see that
the source is strongly absorbed, even though its continuum brightness
temperature is not particularly high. We will use these observations
for a new discussion of the distance to DR 16.

\subsection{Other observations}

Within the frame of the CGPS there are other datasets which
yield additional information on the DR 16 area. We summarize these and
other publicly available data in Table ~\ref{tab-other}. Sadly,
adequate observations of the molecular gas are not available. The work
of Cong (\cite{cong}) and Dobashi et al. (\cite{dob}) and the
collection of CO survey results of Leung \& Thaddeus (\cite{leung}) do
little more than tell us that molecular gas exists in the general DR
16 area. Their lack of resolution and/or coverage prevent examination
of details. 
The NRAO VLA Sky Survey (NVSS) (Condon et al. \cite{nvss}) 
covers the source area. Due to its lower resolution and insensitivity to 
extended structure our VLA map contains more fine structure. 
Near 
infrared (from 2MASS) and optical (POSS) images will be referenced later.

\section{Spectral energy distribution}

To establish the nature of DR 16 we began by deriving radio and
infrared spectral indices. The difficulty was the complexity of the
background emission. For a convenient and uniform approach at all
frequencies we used the `fluxfit' routine from the DRAO Export
Software Package, which fits a Gaussian source component and a
twisted-plane background simultaneously. With the present data this
method was preferable to the alternative of defining a complex
background and integrating the excess emission.  However convenient
the technique, we do not mean to imply that the source consists of
actual Gaussian components. It is clear from Figs.~\ref{kar74} to
\ref{cgpscont21} that the form of the background and the number of
Gaussians will be strongly resolution dependent. At a given frequency
we started with a number of components estimated by eye, and added or
deleted components by using the $\chi ^2$ discriminator calculated by
the routine. The results of this exercise for the available radio data
are collected in Table ~\ref{gaussra}.
\begin{table}[h]
\caption{\label{gaussra} List of sources from gaussian decomposition.
The source notation (column 1) is frequency/component number. 
}
\begin{tabular}{lccrc}
\hline
Source      & $\alpha $ & $\delta $ & flux & observed \\
            & (B1950) &  (B1950) & density & size \\
            & $20^\mathrm{h}\,30^\mathrm{m}$ & $43\degr$ & [mJy] & \\
\hline
151/1   & ~40.1$^\mathrm{s}$ & ~29\arcmin \,33\arcsec & 1584 $\pm $ 375 & 
2.41\arcmin $\times $ 1.36\arcmin \\
151/2   & 49.0 & 27~~56 & 1249 $\pm $ 691 & 
5.79\arcmin $\times $ 1.20\arcmin \\
327/1   & 33.1 & 30~~13 & 759 $\pm $ 32 & 
2.62\arcmin $\times $ 1.50\arcmin \\
327/2   & 33.9 & 33~~01 & 299 $\pm $ 31 & 
2.50\arcmin $\times $ 1.45\arcmin \\
327/3   & 42.9 & 31~~03 & 1322 $\pm $ 41 & 
3.04\arcmin $\times $ 2.00\arcmin \\
408/1   & 37.6 & 30~~58 & 3232 $\pm $ 327 & 
5.50\arcmin $\times $ 4.04\arcmin \\
408/2   & 48.8 & 33~~33 & 1538 $\pm $ 468 & 
11.6\arcmin $\times $ 3.72\arcmin \\
1420/1  & 31.6 & 32~~16 & 278 $\pm $ 18 & 
2.82\arcmin $\times $ 1.18\arcmin \\
1420/2  & 32.6 & 30~~19 & 844 $\pm $ 19 & 
2.52\arcmin $\times $ 1.51\arcmin \\
1420/3  & 42.3 & 31~~09 & 2105 $\pm $ 28 & 
3.64\arcmin $\times $ 2.17\arcmin \\
1515/1  & 31.1 & 31~~31 & 78 $\pm $ 5 & 
1.38\arcmin $\times $ 0.42\arcmin \\
1515/2  & 31.7 & 30~~28 & 1078 $\pm $ 10 & 
2.43\arcmin $\times $ 1.28\arcmin \\
1515/3  & 34.5 & 29~~52 & 19 $\pm $ 3 & 
0.56\arcmin $\times $ 0.31\arcmin \\
1515/4  & 37.3 & 32~~45 & 1202 $\pm $ 16 & 
2.92\arcmin $\times $ 2.40\arcmin \\
1515/5  & 38.3 & 30~~41 & 91 $\pm $ 8 & 
0.67\arcmin $\times $ 0.40\arcmin \\
1515/6  & 41.9 & 31~~36 & 165 $\pm $ 14 & 
1.34\arcmin $\times $ 0.66\arcmin \\
1515/7  & 43.5 & 30~~14 & 1812 $\pm $ 17 & 
3.25\arcmin $\times $ 2.57\arcmin \\
1515/8  & 44.8 & 31~~02 & 123 $\pm $ 13 & 
0.98\arcmin $\times $ 0.86\arcmin \\
1515/9  & 47.0 & 31~~32 & 42 $\pm $ 5 & 
1.74\arcmin $\times $ 0.33\arcmin \\
4800/1  & 38.9 & 31~~03 & 3805 $\pm $ 210 & 
4.35\arcmin $\times $ 4.08\arcmin \\
\hline
\end{tabular}
\end{table}
The sum of the components at each frequency should be a good measure
of the total flux density of DR 16. However, the spectrum constructed
from these sums was rather ragged. We supect that the combination of
angular resolution and frequency dependent background effects are to
blame. For these reasons we excluded components 151/2, 408/2, and
1515/4.
\begin{figure}[h]
\resizebox{\hsize}{!}{\includegraphics{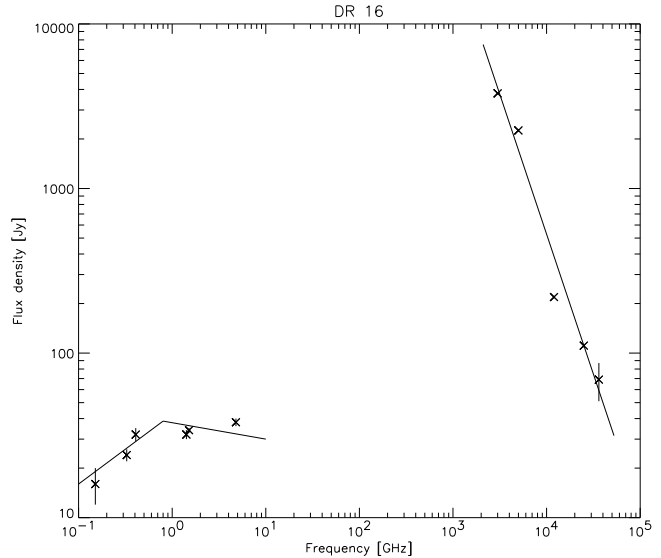}}
\caption{Radio and infrared spectral energy distribution of DR 16.
The radio flux densities have been shifted upwards by a factor of 10.
The broken line through the radio flux densities is for illustrative
purposes; the high frequency part has a slope of $-$0.1.  The gradient
in the IR is $-$1.7 and is also shown for illustrative purposes. }
\label{index} 
\end{figure}
\begin{figure*}
\resizebox{\hsize}{!}{\includegraphics{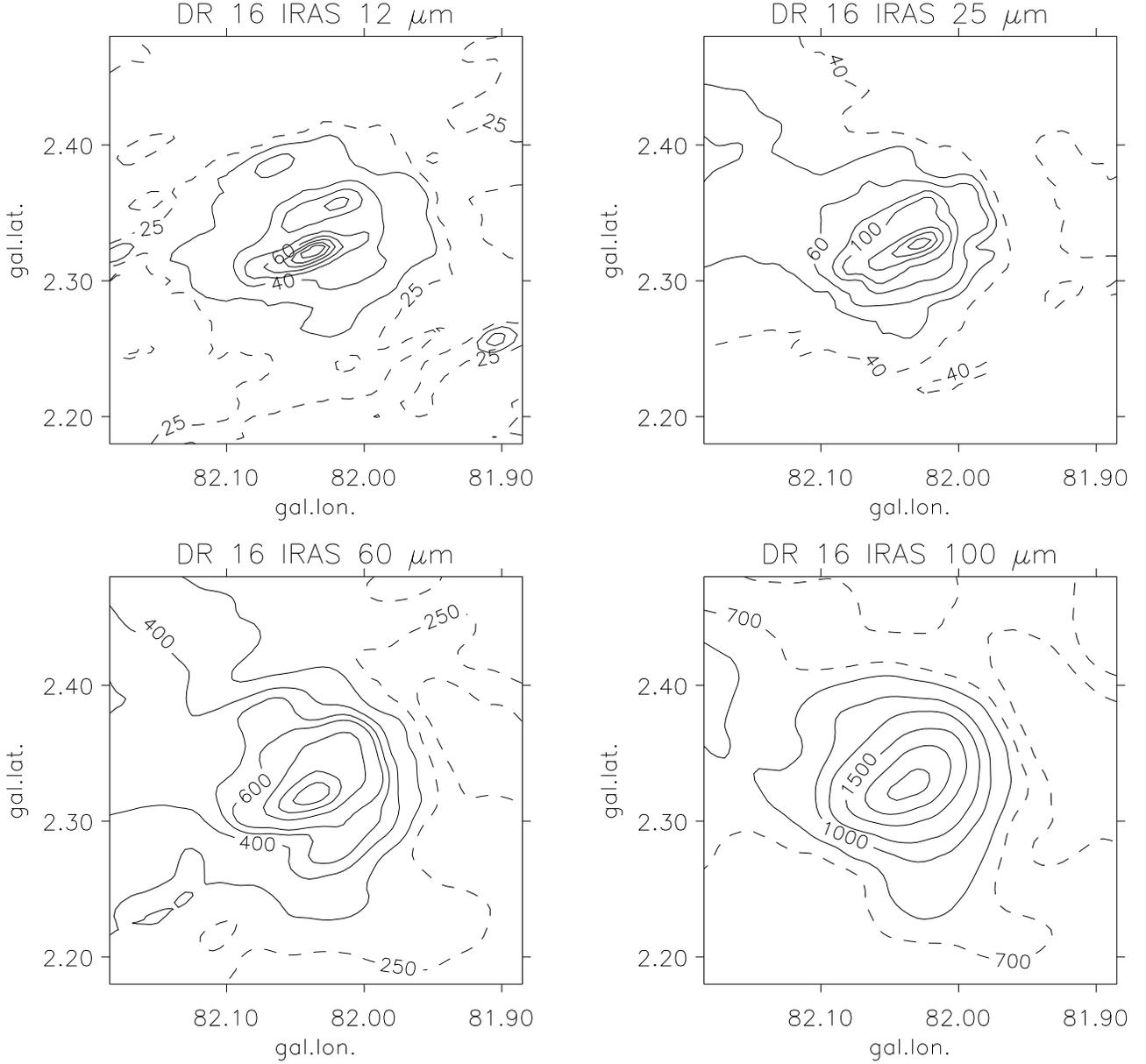}}
\caption{Contour maps of the 4 infrared passbands from IRAS.
Contour units in MJy/ster are: (12$\mu $m) 20, 25, 30, 40, 50, 60, 70,
80, 90; (25$\mu $m) 30, 40, 50, 60, 80, 100, 150, 200, 250; (60$\mu $m) 
200, 250, 300, 400, 500, 600, 900, 1200, 1500; (100$\mu $m) 600, 700, 
800, 1000, 1200, 1500, 1800, 2100. 
Dashed contours indicate a first approximation to the background. 
} 
\label{iras4}
\end{figure*}
The resulting total flux densities of DR 16 proper are plotted in
Fig.~\ref{index}.  The firm conclusion from this exercise is that the
radio emission of DR 16 is optically thin thermal emission above 1 GHz
with a gradual transition to higher optical depth at lower
frequencies. The optical depth of most parts exceeds unity at 151 MHz.

The set of contour maps in the four IRAS passbands in Fig.~\ref{iras4}
reveals a general absence of structure in the source in the infrared.
The background is relatively faint and smooth, and we used the
`fluxfit' routine on these data, just as at radio wavelengths. The
resulting flux densities, together with one obtained in the same way
from the 8.28 $\mu m$ MSX data, are plotted in Fig.~\ref{index}.

There is obviously a non-negligible amount of cold dust at the
location of DR 16. There is no indication that the IR flux density
turns down at 100 $\mu $m towards the Planck maximum, implying a dust
temperature below 25 K, but an actual value cannot be derived; we will
assign $T_\mathrm{dust}\,\approx 20$ K at a later point.  Dust within
\ion{H}{ii} regions is substantially warmer ($\sim 40$ K), and
consequently the radio and infrared parts of the spectral energy
distribution must represent two quite different environments.

\section{21 cm continuum absorption}

The CGPS \ion{H}{i} data cube contains no emission features 
clearly associated with DR 16. On the other hand it is quite obvious that 
there is absorption of continuum emission by local gas to be used for 
foreground and distance discussions.
The optical depth, $\tau $, and the spin temperature, $T_\mathrm{S}$, of
the absorbing neutral hydrogen column can be derived.
For a discussion of the problems and of the
equations we refer to Wendker \& Wrigge (\cite{cx20}, Paper XX). 

The optical depth is connected to three observed quantities via \\ 
$\tau (RV)\,=\,-\ln (1\,-\,(T_\mathrm{b,off}(RV)\,-\,T_\mathrm{b,on}
(RV))/T_\mathrm{c})$ \\ where $T_\mathrm{b,off}$, $T_\mathrm{b,on}$
and $T_\mathrm{c}$ are the expected emission line brightness
temperature, the directly measured line temperature and the continuum
brightness temperature, respectively. The variable RV is radial
velocity referred to the local standard of rest (LSR). DR 16 is
partially resolved in both continuum and line images. We ignore 
the slight difference in angular resolution
between the CGPS continuum and line mosaics (Table~\ref{tab-drao}).
\begin{figure}[h]
\resizebox{\hsize}{!}{\includegraphics{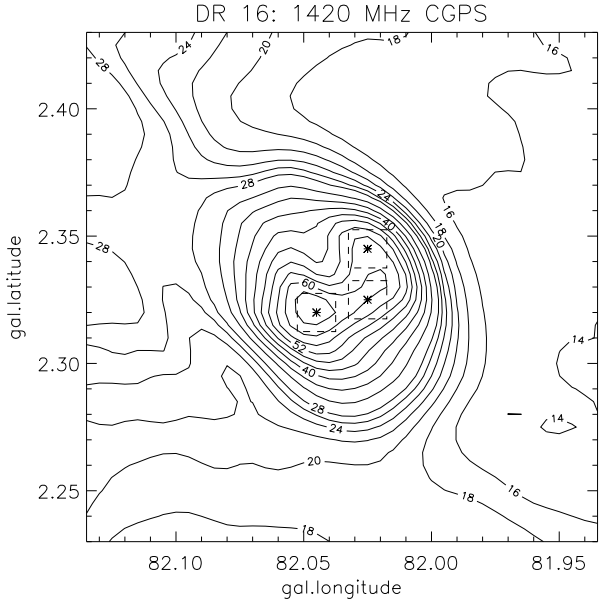}}
\caption{Contour map of DR 16 from the 1420 MHz CGPS
continuum data. Contours range from 14 K to 28 K in steps of 2 K
and from 28 K to 64 K in steps of 4 K. The dashed squares
enclose the $3\,\times \,3$ pixels used for averaging in the 
continuum absorption discussion (centred at the asterisks).
The outer quite regular elliptical contours are used to define the 
geometric centre as $(l,b)=(82.035\degr ,2.325\degr )$.
 } 
\label{dr16c21}
\end{figure}
\begin{figure}[h]
\resizebox{\hsize}{!}{\includegraphics{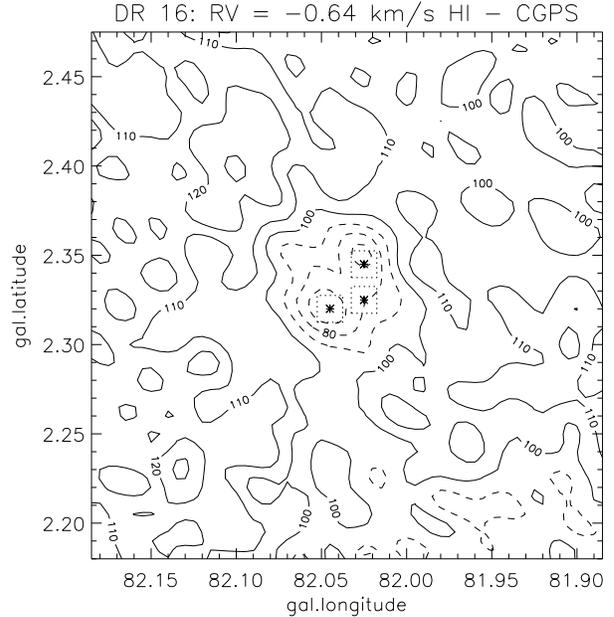}}
\caption{Brightness temperature map of the three-channel average centred on 
$RV\,=\,-0.64$ km/s. The area is somewhat
larger than in Fig. ~\ref{dr16c21} to show that the assumption of   
smooth emission on large scales is justified. A 30 K 
depression due to continuum absorption can be recognized (dashed contours).
The dotted squares and asterisks have the same meaning as in 
Fig.~\ref{dr16c21}.   
 } 
\label{dr16ch72}
\end{figure}
\begin{figure}[h]
\resizebox{\hsize}{!}{\includegraphics{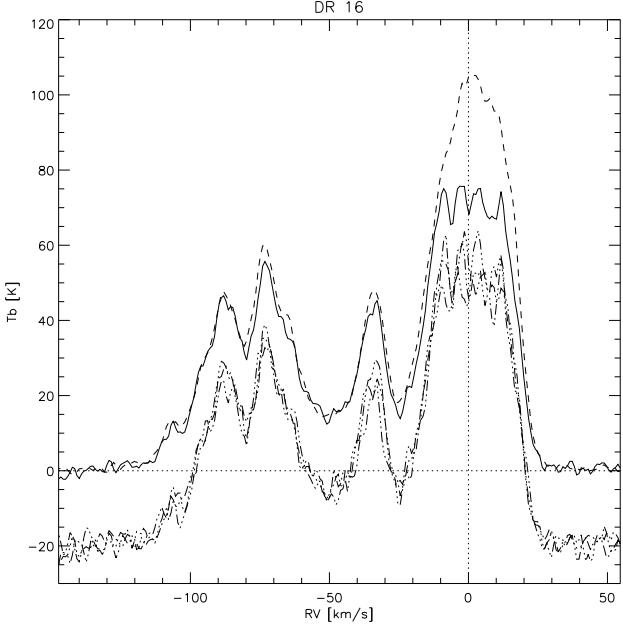}}
\caption{The selected off-sepctrum is plotted as
a dashed curve; it is the average of 72 spectra from points 
in a ring between 10 and 11 pixels away from the geometric
centre. The three different 3 x 3 pixel averages (on-spectra --
dash-dotted) are shifted downward by 20 K for clarity. Their
average is indicated by an (unshifted) solid line.  
 } 
\label{onspec}
\end{figure}
\begin{figure}[h]
\resizebox{\hsize}{!}{\includegraphics{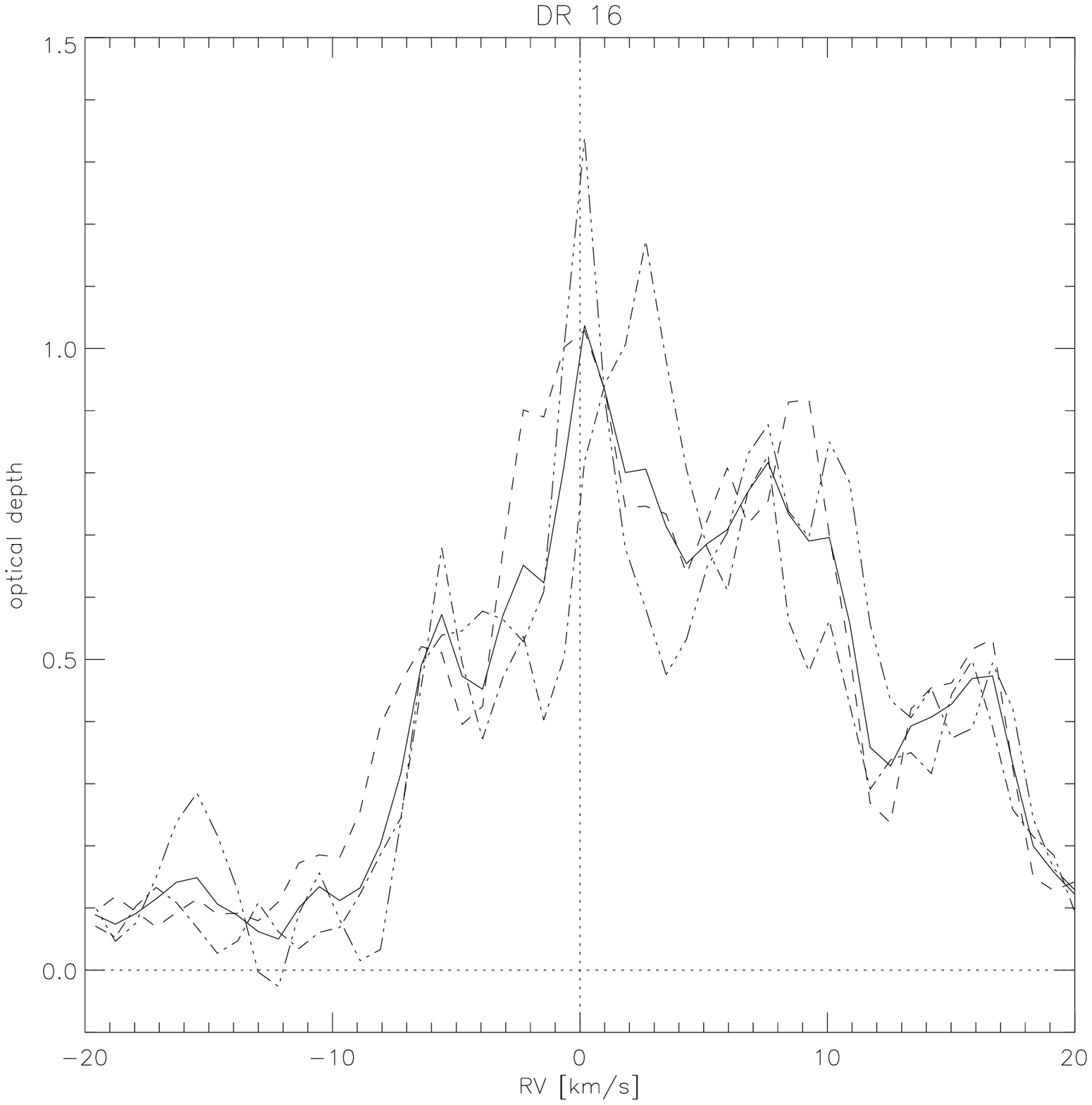}}
\caption{Optical depth versus the radial velocity
interval for the local gas. The three dashed-dotted lines are for the
individual areas and the solid line represents their mean.
 } 
\label{optdep}
\end{figure}

A detailed continuum map is shown in Fig. ~\ref{dr16c21} where three
on-source areas are marked. Nine-pixel averages of these areas are
(from bottom to top): $T_\mathrm{c1}\,=\,65.0$ K,
$T_\mathrm{c2}\,=\,59.0$ K, $T_\mathrm{c3}\,=\,56.7$ K. The rms
scatter of these averages is $\sim$2 K, but the greatest uncertainty
arises from separation into foreground emission which is not absorbed
and source emission which is. Other than DR 16 itself there are at
least three contributions to consider: a remnant brightness
(about 5 K) from the extended ridge (which constituted the tail of the
supposed cometary nebula), a diffuse Galactic component, and the
cosmic microwave background (CMB). Although we will subsequently argue
that the `tail' is not generated by the `head', we nevertheless believe
it is plausibly at the same distance. Continuum emission from the tail
and from the CMB will therefore not confuse the absorption
calculations. From the data of Fig.~\ref{cgpscont21} we estimate that
the diffuse Galactic component could be as large as 12 K, which could
be entirely in the foreground. However, it is more reasonable to
assume that it is spread uniformly along a 10 kpc line of sight
through the disk. For a 
distance of 3 kpc to DR 16, the foreground
contribution would be only 3.6 K. Another way of estimating the
foreground is to associate it with the faint extended \ion{H}{ii}
region connected with the \object{Cyg OB2} association, as discussed
by Huchtmeier \& Wendker (\cite{cx9}, Paper IX). 
Their 11-cm brightness of $\sim$0.8 K translates to 3.2 K at 1420 MHz.
Considering the vastly different
assumptions, the two estimates agree quite well. We will subtract 3 K
from the continuum brightness temperatures mentioned above, note that
the uncertainty in this number will subsequently dominate the absolute
error budget, and will formally use a 4\% error for $T_\mathrm{c}$.

The channel maps show that the scale of the \ion{H}{i} emission is
larger than the continuum source, with emission features on average
$>10\arcmin $. A typical channel map is shown in Fig.~\ref{dr16ch72}.
From inspection of this figure and Fig.~\ref{dr16c21}, the off-spectrum
should be 
derived beyond the 100 K line contour and below the 30 K continuum 
contour.
Trials with various ring sizes
showed hardly any difference between off-spectra obtained at different
radii, and for convenience we chose the one closest to the source.
The on-source spectra for the three areas and the off-source spectrum
are shown in Fig.~\ref{onspec}.

We now obtain optical depths and spin temperatures.
Fig.~\ref{optdep} shows 4 curves for $\tau $: the 
results for
the individual areas and their mean (solid line). The error budget in
the continuum brightness temperatures is dominated by the separation
of foreground from background, and we estimate this to be $\sim$4\%
(which may be optimistic). The uncertainty in on-source line
temperatures comes from noise (receiver plus sky) and variation in
emission gradients, and we formally set this to 5 K for a signal of
$\sim$70 K, or 7\%. The uncertainty in the off-source spectrum is at
least the scatter around the average, and we assign a further 7\% to
this. Conservatively adding these contributions linearly brings us
close to 20\%, which translates to an uncertainty in optical depth of
0.2.  We therefore regard all values of ${\tau}<{0.2}$ as not
meaningfully measured.  Looking at the scatter between the three
individual areas we see no significant foreground fluctuations over DR
16 and find very good agreement between the areas. We therefore use
the averaged curve to derive $\tau $ and $T_S$.

First we interpret these results in velocity space. Continuum
absorption becomes noticeable around +19 km/s where the emission rises
to the point where $T_\mathrm{b}\,>\,30$ K. $\tau$ reaches a first
maximum of nearly 0.5 at +16.5 km/s, with $T_\mathrm{S}$ just below
200 K. This \ion{H}{i} must be the closest neutral layer in the Cygnus
Rift.  $\tau$ drops slightly at +12 km/s with an increase in
$T_\mathrm{S}$ to around 300 K, possibly corresponding to a neutral
but warm outer layer of the Rift (there will be some overlap in
velocity space of gas at distinctly different distances, and this
could dilute the changes in $\tau $ and $T_\mathrm{S}$). After
dropping back to 200 K, $T_\mathrm{S}$ does not vary significantly out
to $-$5 km/s, although different absorbing structures can be
identified by the variations in $\tau $.  This faint anticorrelation
between $\tau $ and $T_\mathrm{S}$ plausibly fits into the standard
picture where higher optical depth means higher volume density, which
in turn implies lower temperature because of greater cooling
effeciency. The peak in $\tau $ at 0 km/s 
probably marks the
densest part of the local Rift gas.  $\tau $ reaches a final peak at
$-$6 km/s, after which it quickly drops to unreliable levels. We
intepret this as the last layer of neutral hydrogen in front of DR 16,
which we place at $-$7 km/s in radial velocity space.

We also computed absorption spectra for the two continuum maxima of the 
tail. Absorption is detectable at both maxima 
with roughly similar
optical depths, but the scatter rises dramatically because the
continuum temperature is much lower ($\sim$35 K) and off-spectra must
be derived from regions parallel to the ridge but quite far removed.
Nevertheless, the results are very consistent with those for DR 16
proper, and from the fact that continuum absorption is visible out to
$-$6 km/s we conclude that the tail must be at the same distance.

The column density of neutral hydrogen can be deduced by summing the
relevant channels of the corrected emission line profile up to the
continuum source. Using the standard relation, a channel contains
$N_\mathrm{HI}\,=\,1.823\cdot 10^{18}\cdot T_\mathrm{S}(RV)\cdot \tau
(RV)\cdot \Delta RV$ where $\Delta RV\,=\,0.82446$ is the channel
spacing. We obtain $N_\mathrm{HI}\,=\,6\cdot 10^{21}\,
\mathrm{cm}^{-2}$.  Using another standard relation
($N_\mathrm{HI}\,=\,5.9\cdot 10^{21}\cdot E_\mathrm{B-V}$) we derive a
total visual extinction of $3^\mathrm{m}$ towards 
DR 16 which is about half the value deduced in Paper V. We suspect that a
substantial contribution to the extinction not represented by
neutral hydrogen is connected to two
molecular clouds which were seen in H$_2$CO (Paper XVII) at $-$0.7 and
+3.4 km/s where the above relation does not hold. Piepenbrink \&
Wendker (Paper XVII) estimate that the integrated effect of the
H$_2$CO containing clouds represents more than $2^\mathrm{m}$ of
extinction, which would close the gap.

\section{Near infrared and optical data}

\subsection{MSX}

\begin{figure}
\resizebox{\hsize}{!}{\includegraphics{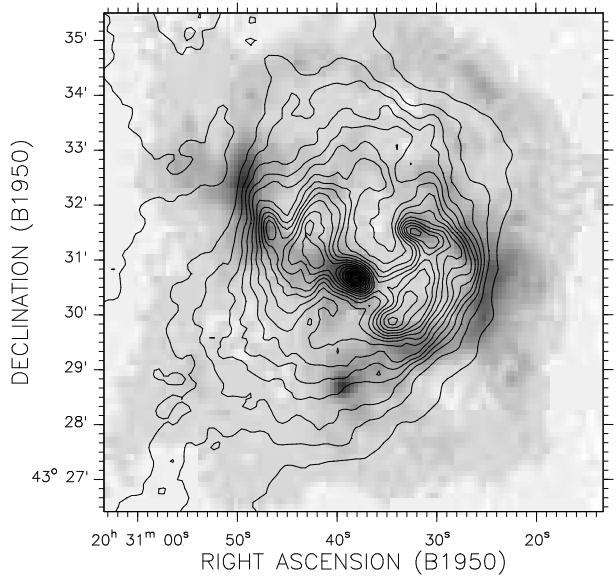}}
\caption{Grayscale image of the MSX 8.28 $\mu $m map overlaid with
radio contours from the VLA observations from Fig.~\ref{vladr16}. The
infrared brightness increases from white to black.} 
\label{ir8radio}
\end{figure}

In discussing the spectral energy distribution we remarked that the
IRAS maps do not give us much information on the dust distribution.
The spectral index derived in that section implies that there is a
large amount of cold dust, but also some warmer dust. We cannot tell
whether the warm and cold components are mixed (usually not the case)
or the location of the warm dust. This can be remedied to some extent
by looking at the MSX data (referenced in Table~\ref{tab-other}). The
absolute calibration of the MSX data is somehat uncertain, and the
good fit of the MSX flux density with the IRAS spectrum may be a
coincidence. However, the MSX image is quite useful. If we consider
the wavelength of 8.28 $\mu $m to represent predominantly warm dust
(say 40 K), then Fig.~\ref{ir8radio} shows that, besides a compact
concentration coinciding with the central radio point source,
filaments of warm dust lie just outside the radio filaments.
The picture which immediately comes to mind is one in which the
radio structures (ionized hydrogen) are photo-dissociation regions
on the edges of dust clouds. We will return to this
interpretation later.

The MSX Archive Point-Source Catalogue contains nine entries over the
area enclosed by the low contours of the MSX and VLA images. Seven of
them coincide with extended structures, and are probably not real
point sources, although very faint point sources might be buried
in the extended emission. The two remaining entries
(\object{MSX5C G082.0329+02.3247} and \object{MSX5C G082.0087+02.3024},
later referred to as MSX8-1 and MSX8-2) appear to be pointlike.
MSX8-2 has an optical counterpart in both the blue and red POSS
images, and we regard it as a foreground star; its high brightness
in the near-infrared photometry of the 2MASS survey argues for a very
late type star. MSX8-1 seems to have a core-halo structure, and coincides
with the central VLA source (1515/5 in Table~\ref{gaussra}). We discuss
its nature and identification in a later section.

\subsection{J, H, K photometry}

\begin{figure}[h]
\resizebox{\hsize}{!}{\includegraphics{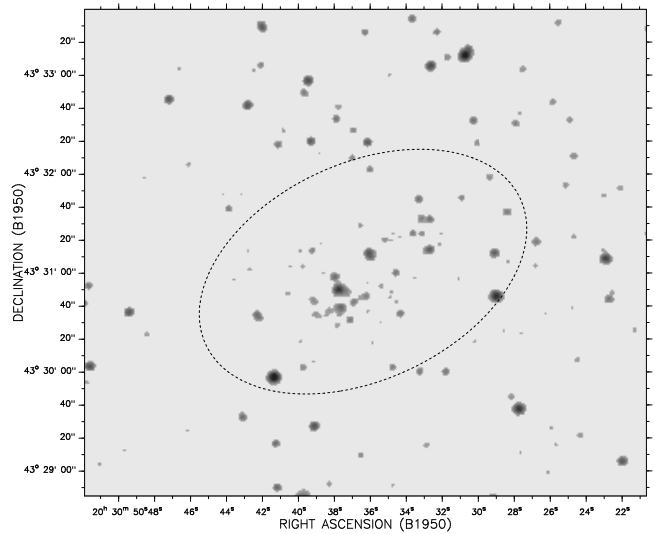}}
\caption{$\mathrm{K_s}$ band image of the DR 16 area from the 2MASS
data.  The ellipse encloses the area which is considered the extent of
the cluster. }
\label{ksgrau}
\end{figure}

Two (observationally) different surveys of the DR 16 area in the
standard near-infrared bands (J, H, K) are available. Comer\'on \&
Torra (\cite{com}) observed at the Observatorio del Teide and Dutra \&
Bica (\cite{dutra}) extracted a dataset from the 2MASS data base.
Both concluded from the images that a stellar over-density is present
at the DR 16 radio position and derived colour-magnitude diagrams
(CMDs) and distances for the central parts of the proposed clusters.
In view of the rather dissimilar distance estimates (1.8 kpc and 2.7
kpc) we obtained and examined both data sets.  Plotting both
photometry and astrometric positions it became apparent that the
photometric values do agree within the mutual errors very
satisfactorily.  The main differences are, first, a somewhat fainter
catalogue limit for the 2MASS data, and, second, the assignment of a
quite different area to the new cluster (in each case determined by
eye). For the first reason we decided to use the 2MASS data to have an
independent third look. A greyscale image of the cluster area is given
in Fig.~\ref{ksgrau} for illustrative purposes.

We first produced a map of the stellar density distribution in the
K$_\mathrm{s}$ band (the band least affected by extinction) by
counting objects in 0.75 arcmin$^2$ circles. The central density
enhancement was well fitted by a Gaussian, giving a geometric centre
of (B1950: $20^\mathrm{h}30^\mathrm{m}36.4^\mathrm{s}, \,43\degr
31\arcmin 01\arcsec$) and a (Gaussian) size of $1.6\arcmin \times
1.1\arcmin$ at a position angle of $115\degr $. (We do not imply that
the real distribution is Gaussian.) The central position lies in the
middle of a small hole, whose significance is not obvious (statistical
fluctuation, additional patch of extinction, or division into two
sub-structures come to mind). The centre is about 50\arcsec\
north-west of the central VLA source (1515/5).

\begin{figure}[h]
\resizebox{\hsize}{!}{\includegraphics{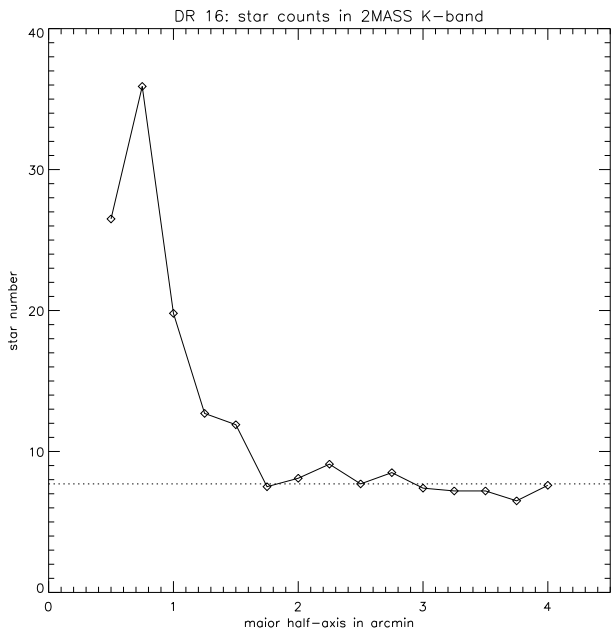}}
\caption{Stellar density in stars per arcmin$^2$ plotted against
the length of the major half-axis of elliptical rings with an axial ratio 
of 1.6:1 for the K$_\mathrm{s}$ band. The adopted background level of 
$7.7\pm 0.7$ stars per arcmin$^2$ is the average of all values beyond
1.75\arcmin\ and is indicated by a dotted line. The 
geometric centre is given in the text. } 
\label{nvsr}
\end{figure}

Fig.~\ref{nvsr} shows stellar density within elliptical rings around
the cluster centre. The rings have a width of 0.25\arcmin, and an
axial ratio of 1.6, with the major axis at a position angle of
115\degr. The existence of a star cluster above a background of about
8 stars per arcmin$^2$ is beyond doubt. The peak is off centre because
of the hole already mentioned.  The total extent is $3.5\arcmin $
along the major axis.  The ellipticity may be more pronounced in the
outer parts than we have assumed, as may be seen in Fig.~\ref{ksgrau}.
Subtracting a background level of 7.7 stars per arcmin$^2$ from the
total of 93 stars inside this ellipse (6.01 arcmin$^2$) leaves about
47 stars in the cluster.  Only 6 stars are needed to fill the central
depression, and the stars in the second ring appear to be the most
heavily reddened. The central minimum is then probably the result of
additional extinction, but the existence of two substructures cannot
be ruled out.

The placement of a reddened zero-age main sequence on a cluster 
in a near infrared CMD is highly ambigious.
The reddening vector is nearly parallel to the colour axis in a 
$m_\mathrm{K_s}$ vs. ($\mathrm{J-K_s}$) CMD for the upper main sequence, 
and the distance modulus is parallel to the apparent magnitude axis. 
Dutra \& Bica (\cite{dutra}) used a template method which assumes that for a
small young cluster the tenth brightest star has an absolute magnitude 
close to $0^\mathrm{m}$ (spectral type about A0). The template cluster is
\object{NGC 6910} from the optical photometry by Delgado \& Alfaro 
(\cite{del}). However, for the  ninth brightest star of their member list we
infer from the BDA database (Mermilliod \cite{merm}) a spectral type 
of B1V. Therefore the Dutra \& Bica (\cite{dutra}) assumption does not 
seem to be well justified. 

\begin{figure}[h]
\resizebox{\hsize}{!}{\includegraphics{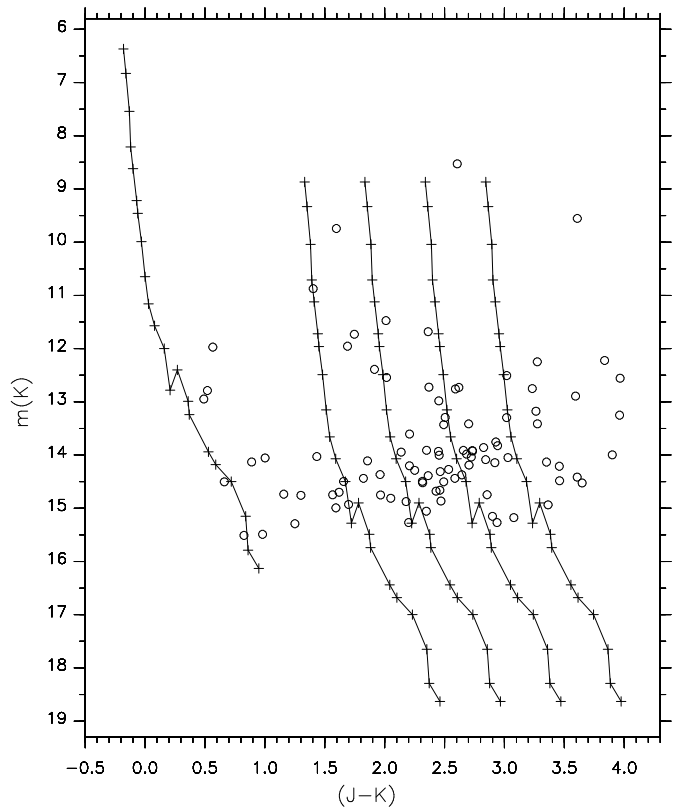}}
\caption {The $m_\mathrm{K}$ vs. ($\mathrm{J-K}$) CMD of all objects
inside the elliptical area assigned to the DR 16 star cluster. The
most leftmost line is the theoretical unreddened zero-age
main sequence shifted to a distance modulus of 10$^\mathrm{m}$, spanning 
spectral types O9V to M5V. The other lines represent the same sequence 
shifted to a distance modulus of 12.5$^\mathrm{m}$ and successively 
reddened by 9$^\mathrm{m}$, 12$^\mathrm{m}$, 15$^\mathrm{m}$ and 
18$^\mathrm{m}$ of visual extinction A$_\mathrm{V}$. } 
\label{cmd}
\end{figure}

A theoretical unreddened zero-age main sequence was read from the
tables given by Cox (\cite{cox}) for the $m_\mathrm{K}$
vs. ($\mathrm{J-K})$ CMD. The tables have limited coverage of
spectral range, only from O9 to M5, but seem to be the best
available. From the parametrization of the extinction law by Cardelli
et al. (\cite{car}) and values given by Mathis (\cite{mat}) we
calculated the reddening vector to be $E_\mathrm{J-K}\,=\,0.168\cdot
A_\mathrm{V}$ where $A_\mathrm{V}$ is the total visual extinction to
be used. The unreddened zero-age main sequence shifted to a distance
modulus of 10$^\mathrm{m}$ is plotted in Fig.~\ref{cmd}. It will give
some indication which stars may be unreddened foreground objects and we
see that three of the brighter stars belong to this group. As they are
also visible on the POSS plates with optical colours $B-V\,\sim
\,1.5^\mathrm{m}$ they are indeed late type foreground stars.
For the brighter stars there is quite a gap in colour between the
unreddened sequence and the suspected cluster stars. The minimum
reddening is then 1.4$^\mathrm{m}$ in $\mathrm{J-K}$, corresponding
to about 8$^\mathrm{m}$ of A$_\mathrm{V}$. The scatter for stars with
a redder colour indicates that substantial cluster and/or
circumstellar reddening occurs. In the absence of knowledge of the
spectral type of any of the stars it seems impossible to `fix' a
number so that a distance can be inferred.  The only conclusion
permitted by the CMD is that the data are consistent with any distance
between 1 and 4 kpc. We will illustrate this with one numerical
exercise.  Assume a distance modulus of 12.5$^\mathrm{m}$ (3 kpc); the
star \object{2MASS J2032214+434107} with
$m(\mathrm{K_s})=9.74^\mathrm{m}$ and (J-K$_\mathrm{s}$)=1.60 then has
$E_\mathrm{J-K_s}=1.76^\mathrm{m}$ corresponding to
$A_\mathrm{V}=10.5^\mathrm{m}$. Shifting it to the unreddened sequence
leads to $M_\mathrm{K_s}=-3.9^\mathrm{m}$ which in turn corresponds to
a spectral type of O9V. Its extinction therefore occurs mostly in the
foreground, and the additional 1$^\mathrm{m}$ of extinction could
either be circumstellar or local to DR 16.  For such a star the
expected apparent visual magnitude is about 18.4$^\mathrm{m}$.  But
the star is visible on the red POSS plate with a magnitude around
17$^\mathrm{m}$ (GSC) and is just visible on the blue plate! And if we
shift the distance modulus by 1$^\mathrm{m}$ in either direction we
still obtain equally plausible and meaningful numbers. This exercise
can be carried through for all stars above $m(\mathrm{K_s})\sim
13.5^\mathrm{m}$, of which the vast majority are not, and should not be,
visible on the POSS plates, confirming that they are highly reddened
stars, not intrinsically faint ones. Below this value a distinction
between slightly reddened foreground and highly reddened cluster stars
can no longer be made, which explains the widely scattered band at the
bottom of the CMD. The same arguments make it highly likely that at
least the brighter stars in the area are cluster members and not field
stars. Picking some sort of mean reddening from the CMD is not easy.

In the end, the star cluster is only well defined by the projected
number density distribution. Its visual foreground extinction is quite
large (8$^\mathrm{m}$ to 9$^\mathrm{m}$) and is most probably produced
by the Cygnus Rift at a distance of only 600 pc, although a distance
of around 3 kpc or more is supported by the CMD.

\section{On the distance of DR 16}

Two different approaches are available to determine the distance to
the DR 16 complex. On the one hand we can use spectral-line data for a
kinematic distance. Three published recombination line observations
are available, $-$7 km/s from H110$\alpha $ (Paper XVII) for DR 16
proper, $-$5.2 km/s (also Paper XVII) for the tail, and $-$5.1 km/s
from H87$\alpha $ by Lockman (\cite{lock}) but with an angular
resolution that enclosed DR 16 and at least part of the tail. From our
continuum absorption study we now add $-$5 to $-$7 km/s as a very
well determined distance in radial velocity space. Using the rotation
curve of Fich et al. (\cite{fich}) this leads to a distance range of
2.5 to 3.5 kpc. Although kinematic distances are notoriously uncertain
at the Galactic latitudes under consideration, DR 16 seems firmly
placed behind the Cyg OB2 association and its related features.

On the other hand we have the two recent attempts to prove the
existence of an open cluster connected to DR 16 using near infrared
photometry (JHK bands) and thereby derive a distance with the help of
a colour-magnitude diagram (CMD). As we showed above, the Dutra \& Bica
(\cite{dutra}) value of 1.8 kpc is very uncertain because the notion
of a mean cluster reddening from the scatter in the bottom part of the
CMD may be too optimistic.  Comer\'on \& Torra (\cite{com}) obtained
2.7 kpc using a different estimator (a sort of Zanstra method).
However, we found that the star with the brightest $m(\mathrm{K_s})$
value, \object{2MASS J2032250+434014} (see Fig.~\ref{cmd}), was not
taken into account because it lies outside the area considered to be
the cluster area. Adding this star into the Comer\'on \& Torra
(\cite{com}) formalism leads to a value of 0.8 kpc, indicating the
fragility of this method. Furthermore, the brightest cluster members,
which contribute most of the ionizing photons, mostly have large
circumstellar extinctions. Clearly the methodological problem of dust
absorbing such photons has not been properly solved for DR 16.

In spite of the diligent efforts in the near infrared the strongest
constraint remains the radial velocity distance of $-$5 km/s. We
maintain that this means that DR 16 is situated behind the Cyg OB2
cluster and is physically unrelated to it. We will use a distance of 3
kpc from this point, a choice which will appear quite plausible after we
have discussed the nature and origin of the emission. However, we note
that the infrared data are even better compatible with a larger distance.

\section{Nature of the objects}

All the data presented in preceding sections of this paper are
compatible with the idea that DR 16 proper is an isolated cluster of
very young stars. The brightest members can be assumed to be a group
of late O to early B stars capable of producing enough Lyman continuum
photons to ionize the observed \ion{H}{ii}. However, the large number
of stars with substantial circumstellar extinction creates a
difficulty. Why are no ultracompact \ion{H}{ii} regions (UCHIIR)
detected?  If any exist they must be so compact that they can be
observed only at higher frequencies. This would then rule out any
substantial leakage of Lyman continuum photons. Furthermore,
UCHIIR should be seen as point sources at
infrared wavelengths (e.g. Wood \& Churchwell \cite{ed}), which is not
the case. The dust distribution seen in the infrared seems to be
rather diffuse, except for the warm dust, which may be concentrated in
filaments on the outskirts.

The ease with which the VLA map can be decomposed into Gaussian
components could indicate that there are multiple ionization
centres. Indeed, each small component needs only a middle B type star
and the large diffuse emission a single O9V one. However, we would
then expect to see these stars more or less centred on the small radio
components.  The fact that only one such identification appears from a
careful overlay, namely VLA 1515/5 with \object{2MASS J2032213+434056}
(which we refer to together as \#5) argues against this
geometry. We suggest that the radio structures seen are density
fluctuations, with the execption of \#5.

A first estimate of the number of Lyman continuum photons required for
the ionization of the \ion{H}{ii} region DR 16 can be made using
\\
$N_{\mathrm L_c}\,=\,4.761\cdot 10^{48}\cdot T_{\mathrm
e}^{-0.45}\cdot \nu^{0.1}\cdot S(\nu )\cdot
D^2\,\,\mathrm{s}^{-1}$. \\ 
This relation (Mezger \cite{pgm}) is valid only for an optical depth
less than unity, which appears to be the case. Using an electron
temperature of $10^4$ K, an average frequency of 1.45 GHz, a total
observed flux density of 3.3 Jy, and a distance of 3 kpc we obtain
${N_{\mathrm L_c}}=2.3\cdot 10^{48}$ s$^{-1}$. Since we will
subsequently suggest that the three Gaussian components, \#5 (the
central VLA source), 1515/3, and 1515/6, which define the central
dumbell appearance of DR 16, are a Herbig Ae/Be (HAeBe)
star with a giant bipolar
outflow, this value represents an upper limit. We now inspect the
brightest cluster stars to see whether they can deliver so many Lyman
continuum photons. The two stars brightest in 
apparent m$_\mathrm K$ magnitude (2MASS J2032250+434014
and 2MASS J2032126+434103) are
also heavily reddened by circumstellar material (Fig.~\ref{cmd}). 
This additional extinction amounts to 8.5$^\mathrm m$
and 13$^\mathrm m$, respectively. Although their position in the CMD
makes them late O-type stars, they will not contribute to ionization
of the diffuse \ion{H}{ii}, since we consider that they are still
within their parent molecular cocoons. The absence of
UCHIIR around them supports this suggestion --
the UCHIIR are still too small and our frequency is too low. The
third brightest cluster member is \object{2MASS J2032214+434107},
already used as an example in Sect.5 and shown to be most likely a
B0.5V star. Its Lyman continuum photon production is $8\cdot
10^{47}\,\,{\mathrm s}^{-1}$ (Vacca et al. \cite{vacc}). This star is
marginally adequate, and, if supplemented by the next ranked B stars
(type B2), would account for the ionization of the diffuse
\ion{H}{ii}. The star lies near the centre of the object, making it
plausible that DR 16 is an ionization bounded \ion{H}{ii} region.
Remarkably, increasing the distance modulus to 13$^\mathrm m$ (4 kpc)
inverts problem, providing too many photons (which of course could be
absorbed by dust).

The two next brightest 
stars (2MASS J2032214+434107 and 2MASS J2032197+434129) should
have 4$^\mathrm m$ and 6$^\mathrm
m$ of circumstellar (or local) extinction according to the CMD and are
consequently invisible or barely visible on the POSS.

The seventh brightest (2MASS J2032213+434056) star 
deserves special attention. It is the only
one which coincides with one of the radio components, the small
Gaussian component 1515/5 fitted to the central point source in the VLA map.
It is also the only one to coincide with a near-infrared source (see
Figs.~\ref{vladr16} and \ref{ir8radio}; \#5). Working through the same
exercise as above, ($m_{\mathrm K}\,=\,11.7^{\mathrm m}$, $A_{\mathrm
V}\,=\,10.4^{\mathrm m}$ $\rightarrow $ $M_{\mathrm
K}\,=\,-1.9^{\mathrm m}$ $\rightarrow $ type B2V $M_{\mathrm
V}\,=\,-2.4$) we calculate an expected apparent red magnitude of
$17.8^{\mathrm m}$. This appears to be compatible with the GSC red
brightness of $16.7^{\mathrm m}$, given the roughness of the
calculations.  We conclude that \#5 has not yet reached the main
sequence and thus qualifies as a HAeBe star. In addition
we interpret the coincidence with a small-scale radio source and the
dumbbell shaped structure in the radio map as an indication of a
central source plus jet. 
The radio lobes are then remnants of a giant bipolar outflow. Lower mass
stars in a less photoionizing surrounding often produce Herbig-Haro 
objects (HH) in such circumstances. The projected outflow length on
either side is about 1\arcmin\ or not quite 1 pc. In this
interpretation we assume that the diffuse \ion{H}{ii} and the dumbbell
are projected onto each other without physical interaction at the
present time, but both are features of the star formation event of DR
16.

It is one thing to make this qualitative interpretation, but to make
it more quantitative is quite another. In the following discussion we
take properties and relations for HAeBe stars from the review of
Waters \& Waelkens (\cite{haebe}) and for HH objects from that of
Reipurth \& Bally (\cite{hh}).

The requirements for classification as a HAeBe star are only partially
met.  Although it seems certain that this is a B star, we lack a
suitable spectrum to test the emission line requirement; we need a
high-resolution spectrum of a fairly faint star. On the other hand,
the infrared excess is definitely there. The existence of a point
source in the MSX 8.3 $\mu $m band and its visibility at other MSX
wavelengths (Fig.~\ref{ir8radio}) is a strong indicator. Both the radio 
source and its near-infrared counterpart
are definitely elongated in the direction of the alleged jet structure
but hardly perpendicular to it. This is best measured in the radio
domain by the sizes of the fitted Gaussian: the major axis is
40\arcsec\ and the minor axis less than 20\arcsec . However, some sort
of core-halo geometry is equally compatible with the observations. We
emphasize that the data do not have enough angular resolution to
separate emission from the (extended) atmosphere of the star, from the
remaining natal disk, or from the polar jet. Similarly, we cannot
really say whether a jet connects the central source and the bipolar
plumes.  Bipolar outflows are often detected from their CO emission,
and we have already lamented the lack of adequate observations of this
kind. We 
suggest that the radiation from the emergent star
has mostly ionized the outflow plumes. The warm dust, visible at 8.3
$\mu $m, is positioned at the outside of the \ion{H}{ii} as seen from the star
(vividly portrayed in Fig.~\ref{ir8radio}) and we suggest that this
large surface is the working interface between the outflow and the
remaining cold interstellar medium from the parent molecular cloud.

\section{Remarks on the `tail'}

\ion{H}{i} absorption of the continuum emission from the tail is 
detected albeit very weak out to the same radial velocity as for DR 16
(see Sect. 4). Recombination lines were found at similar velocities 
(Sect.6). Landecker (\cite{tll}) detected a H166${\alpha}$ line with a
0.6\degr\ beam indicating that widespread \ion{H}{ii} with slightly
negative radial valocities must be present. Cong (\cite{cong}) found
a molecular cloud extending 44\arcmin\ by 20\arcmin\ with a radial velocity 
of about $-$2.5 km/s. All these data provide circumstantial evidence
that the tail should lie at about the same distance as DR 16, but none
imply any physical interaction.
 
In the original low-resolution image (Fig.~\ref{kar74}) the
source flares out considerably. It is apparent from
Fig.~\ref{cgpscont21} that this impression is created by long
filaments with varying brightness which do not cross DR 16 implying that
they are not part of it. The western arc is narrow
with a radius of at least 0.5\degr\ ; its outline is lost in the general
confusion in the area. A working hypothesis is
that a shock of distant origin ran into the denser medium of the DR
16 complex, producing the filamentary structure. For the
tail itself, we suggest that it is part of a ring-like feature with a
radius of about 6\arcmin\ (Figs.~\ref{vlaganz} and~\ref{cgpscont21}),
well defined by emission from half of its perimeter, and by an
apparent central hole with a linear extend of about 10 pc. 

More striking than the ring itself is the point source at $\alpha
_{(1950)}\,=\,20^h31^m40.82^s$, $\delta _{1950}\,=\,43\degr 43\arcmin
34.7\arcsec $, very close to its centre. However, the point source
seems to be non-thermal, with flux densities of $117\,\pm \,1$ mJy at
1515 MHz, $113\,\pm \,6$ mJy at 1420 MHz, $249\,\pm \,15$ mJy at 327
MHz and $879\,\pm \,202$ mJy at 151 MHz, indicating a spectral index
of around $-$0.8. This would be compatible with an extragalactic source
such as an AGN or a quasar.

While examining the optical appearance of the area we noticed that the
three stars brightest in m$_\mathrm K$ are all about 1\arcmin\ south
of the centre of the ring (at B1950 positions of 20$^{\mathrm
h}$31$^{\mathrm m}$40.85$^{\mathrm s}$/43\degr 42\arcmin22.0\arcsec,
20$^{\mathrm h}$31$^{\mathrm m}$45.31$^{\mathrm s}$/43\degr
42\arcmin11.7\arcsec, and 20$^{\mathrm h}$31$^{\mathrm
m}$39.35$^{\mathrm s}$/43\degr 41\arcmin43.1\arcsec\ ). Their colours 
indicate heavy reddening. From the magnitudes given in the
USNO-B and 2MASS catalogues they could be O5 to B1 stars at 3 kpc
distance with a reddening of $>6^{\mathrm m}$, just the amount of
foreground reddening for the DR 16 cluster. The ring may have been
created by the action of the stellar winds of these stars on
their surroundings, but we have not proved any connection.

\section{Summary and discussion}

We began this study with the idea that we were seeing an object
which seemed to be a very large cometary 
nebula. In summary the higher resolution data have revealed that we were 
looking at a complex of different objects. The main radio continuum source, 
now identified as DR 16 proper, 
is the common \ion{H}{ii} region of an
open star cluster. The cluster is small, with a total mass of a few
hundred solar masses, and a substantial fraction of the low-mass stars
have yet to reach the main sequence. On the other hand, the fact that
at least some of the O stars have shed their natal envelopes and are
ionizing a common \ion{H}{ii} envelope points to an age of some
hundred thousand years.  However, DR 16 is not an isolated group.  We
have found several signs of interaction with surrounding material. One
of the neighbouring features, in our original picture the `tail' of
the cometary nebula, lies at roughly the same distance and is probably
interacting with the same molecular cloud as DR 16.

We used various methods to estimate the distance to the complex and
derived a best value of 3 kpc. The uncertainty on this number is
probably 0.5 kpc, with the near distance constrained by the properties
of the stellar cluster and the far one by the radial velocity of the
\ion{H}{ii} region. This distance is large enough that physical
interaction between DR 16 and Cyg OB2 seems quite implausible.

Despite having many new observations, the lack of adequate CO data
impedes interpretation. Cong (\cite{cong}) estimated that there is
about 85 000 $M_{\sun }$ of molecular material in the area (scaled by
us for distance). What is required is a fully sampled CO map with a
resolution of 1\arcmin\ or better. We predict that such a map would
show much small-scale structure. 

The internal structure of the DR 16 \ion{H}{ii} region is rather
smooth, with only minor density fluctuations after the components
1515/5, 1515/3 and 1515/6 are subtracted. These three components
together define the structure we have called the dumbbell. We have
shown that enough ionizing photons are available from stars on or near
the main sequence to produce a classical \ion{H}{ii} region. The lack
of UCHIIR around those stars with substantial circumstellar material
could be an age effect. This idea is supported by the recent 
considerations of Keto (\cite{keto}) who has argued that in the beginning
the ionization front will stall at a very short distance from the star 
for a rather long time. 

This leaves the dumbbell as the most interesting individual object.
Furthermore, it has an early type star within its central radio
source.  That the dumbbell is a single physical object is not
necessarily obvious. The best evidence is perhaps in
Fig.~\ref{ir8radio} where the two outer components appear to be
confined by the only two well-defined substructures in the dust
distribution. Furthermore, the central radio source is definitely
elongated in the direction of the lobes. Together these clues suggest
a bipolar outflow (keep in mind that the torus required for the flow
is normal to the dumbbell axis). 

\section{Conclusions}

Part of the following conclusions are based on two review
papers, one on UCHIIR by Churchwell (\cite{edrev}), and one on Herbig
Ae/Be stars by Waters \& Waelkens (\cite{haebe}).  Massive stars
produce UCHIIR which are dispersed after the stars reach the main
sequence. The oldest, most massive stars in the DR 16 cluster have
nearly all reached this state and can now provide the ionization of
the dispersed diffuse gas.  The younger ones are far enough from the
main sequence that their small UCHIIR are not observable.  As the
lines corresponding to stellar evolution and to decreasing
circumstellar extinction are nearly parallel in the CMD we have
available (see Fig.~\ref{cmd}), we cannot provide any more details of
this picture.  The lower mass limit for young stars to which this
picture applies is usually put around 10 $M_{\sun }$ (or spectral type
B0.5). Less massive pre-main-sequence stars are usually called HAeBe
stars and develop more slowly towards the main sequence, but can shine
through their natal cloud much sooner since their accretion disks are
less massive. Probably as a consequence of this, outflows are less
vigorous and hardly observed at stages near the main sequence.  Images
of the thermal radio emission portray the average electron density
along the line of sight, and this can smooth out small-scale
structures. Remains of old UCHIIR and of bipolar outflows, which are
probably ubiquitous in such a region, are thus not visible in our
radio maps.

Our suggestion that the central star of the source 1515/5 is a HAeBe
star and that the \ion{H}{ii} structures 1515/3 and 1515/6 are the
working surfaces of the lobes of a bipolar outflow impinging on
neutral and/or molecular material implies the following specific
evolutionary status.  The star is close enough to the main sequence
that typical HAeBe properties are weak (this statement can be tested
by spectroscopy).  Its mass is large enough that it also has
properties usually ascribed to the massive star picture i.e. an
earlier strong outflow which has now ceased.  On the other hand, the
outflow is sufficiently recent that the density fluctuations it
produced are still visible, although its ionization may by now be
maintained by the diffuse radiation rather than by the originating
star. This transition status makes the star quite interesting.

On the grander scale, DR 16 can now be added to the list of sources in
Cyg X which comprise an \ion{H}{ii} region and a small stellar cluster
and lie at about a distance of 3 kpc. This is beyond the zone of
influence of the Cyg OB2 association, often regarded as all
dominating.  The Cyg OB2 association was suggested to be a young
globular cluster rather than a classical open cluster (Kn\"odlseder
\cite{knoe}).  Distances around 3 kpc have been suggested for
\object{DR 21} and \object{DR 23} (Papers XVII and XI,
respectively). The projected distance of DR 16 from these is only 100
pc or so, which may indicate that this part of the local spiral arm
contains an extended complex of small star forming regions. DR 16
itself is quite bright, about one third of the Orion
nebula.  Compared to DR 15, itself a bright and conspicuous source in
Cyg X, it is about 15 times brighter. On the other hand, it has only
about 1/7th of the radio power of DR 21, usually regarded as a
prominent cluster of just forming stars.

\begin{acknowledgements}

The Canadian Galactic Plane Survey (CGPS) is a Canadian project with 
international partners. The Dominion Radio Astrophysical Observatory 
is operated as a national facility by the National Research Council 
of Canada. The CGPS is supported by a grant from the Natural Sciences 
and Engineering Research Council of Canada.
This research has made use of the NASA/IPAC Infrared Science Archive, 
which is operated by the Jet Propulsion Laboratory, California Institute 
of Technology, under contract with the National Aeronautics and Space 
Administration.
This research has made use of the VizieR catalogue access tool, CDS, 
Strasbourg, France
This publication makes use of data products from the Two Micron All Sky 
Survey, which is a joint project of the University of Massachusetts and 
the Infrared Processing and Analysis Center/California Institute of 
Technology, funded by the National Aeronautics and Space Administration 
and the National Science Foundation.

\end{acknowledgements} 



\end{document}